\documentclass[journal,comsoc]{IEEEtran}

\usepackage[T1]{fontenc}

\usepackage{cite}

\ifCLASSINFOpdf
   \usepackage[pdftex]{graphicx}
\else
   \usepackage[dvips]{graphicx}
   \DeclareGraphicsExtensions{.eps}
\fi
\usepackage{amsmath}
\usepackage{multicol,lipsum}
\usepackage[dvipsnames]{xcolor}
\usepackage[normalem]{ulem}

\DeclareMathOperator{\sinc}{Sinc}
\DeclareMathOperator{\rect}{Rect}

\newtheorem{lemma}{\underline{Lemma}}

\newtheorem{remark}{\underline{Remark}}

\usepackage[cmintegrals]{newtxmath}
\begin{document}
%
\title{Orthogonal Delay-Doppler Division Multiplexing Modulation}
%
%
%
\author{
	Hai~Lin,~\IEEEmembership{Senior Member,~IEEE}, Jinhong~Yuan,~\IEEEmembership{Fellow,~IEEE}%
	\thanks{
  Part of this paper has been presented at the IEEE ICC 2022. }
	\thanks{H. Lin is with the Department of Electrical and Electronic Systems Engineering, Graduate School of Engineering, Osaka Metropolitan University, Sakai, Osaka, Japan (e-mail: hai.lin@ieee.org).}
	\thanks{J. Yuan is with the School of Electrical Engineering and Telecommunications, the University of New South Wales, Sydney, Australia (e-mail: j.yuan@unsw.edu.au).}}
\maketitle

\begin{abstract}
	Inspired by the orthogonal time frequency space (OTFS) modulation, in this paper, we consider designing a multicarrier (MC) modulation on delay-Doppler (DD) plane, to couple the modulated signal with a doubly-selective channel having DD resolutions. A key challenge for the design of DD plane MC modulation is to investigate whether a realizable pulse orthogonal with respect to the DD plane's \emph{fine resolutions} exists or not. To this end, we first indicate that a feasible DD plane MC modulation is essentially a type of staggered multitone modulation. Then, analogous to orthogonal frequency division multiplexing, we propose an orthogonal delay-Doppler division multiplexing (ODDM) modulation, and design the corresponding transmit pulse.
	Furthermore, we prove that the proposed transmit pulse is orthogonal with respect to the DD plane's resolutions and therefore a realizable DD plane orthogonal pulse does exist.
	The orthogonality of this particular pulse significantly eases the derivation of the ODDM's DD domain channel input-output relation, and yields a channel matrix with an elegant block-circulant-like structure. We demonstrate that the ODDM outperforms the OTFS in terms of out-of-band emission and bit error rate, by achieving perfect coupling between the modulated signal and the DD channel.
\end{abstract}

%

\section{Introduction}
%
%
%
%
\IEEEPARstart{O}{ne} of the most important scenarios of the next generation Beyond 5G/6G cellular systems is the extremely high mobility. For example, the Chuo Shinkansen in Japan, a.k.a the Japanese superconducting magnet levitating train between Tokyo and Nagoya scheduled in 2027, will operate at a speed of 500 kilometers per hour. How to achieve a reliable communication in such a high mobility environment is a very challenging issue.
On the other hand, the very crowded microwave band has accelerated the relocation of cellular systems to millimeter wave band, which has huge spectrum vacancy. {The combination of high carrier frequency and high mobility results in severely fast time-varying channels. Obviously, time-variation is one of the key channel features that can be taken into account in the design of next generation waveforms, where a deterministic path-based channel model in the delay-Doppler (DD) domain may provide more insights than the traditional statistical channel model in the time-frequency (TF) domain\cite{fwc}.}

It is well-known that orthogonal frequency division multiplexing (OFDM) modulation, which has been adopted in the current fourth-generation (4G) and fifth-generation (5G) cellular systems, may not work well in a future high mobility environment\cite{ofdm}.
Beside the path gain, each path also introduces a Doppler and a propagation delay, which correspond to the channel's time and frequency domain dispersions, respectively. Thus, the channel in such an environment is both time and frequency selective. After the OFDM signal passing through a doubly-selective channel, the received signal waveform is a superposition of multiple OFDM waveform copies altered by different gains, delays and Dopplers.
Although the OFDM modulation can handle the delay-induced inter-symbol-interference (ISI) by prepending a cyclic prefix (CP) for each OFDM symbol, the remained Doppler-induced inter-carrier-interference (ICI) can cause severe performance degradation.
Usually, the length of one OFDM frame is chosen to let the time-varying multi-path channel be  approximated as time-invariant. In other words, once the frame is contained within the channel's coherent time where the channel is quasi-time-invariant and the Doppler-induced channel variation can be neglected, the received OFDM signal can be demodulated using the discrete Fourier transform (DFT) and a subsequent subcarrier-wise one-tap frequency domain equalizer. Therefore, the classic OFDM modulation can benefit from frequency diversity via coding over subcarriers but does not exploit time diversity.

In contrast to the OFDM's TF plane\footnote{Throughout the paper, we use TF (or DD) plane and TF (or DD) domain interchangeably, as they both were widely used in the literature of OTFS and MC modulation.} modulation, the recently proposed orthogonal time frequency space (OTFS) modulation \cite{Hadani17,hadani2018orthogonal,hadani2018otfs} suggests to modulate information-bearing symbols on the DD plane. By considering the channel's dispersive effects as potential time and frequency diversities rather than undesired impairments, the OTFS modulation shows superior performance over the classic OFDM modulation in high mobility environment.
The rationale behind the superiority of OTFS modulation is the natural physical representation of the doubly-selective channel in the DD domain, namely the DD channel impulse response, where each path is represented as a quasi-time-invariant channel tap on the DD plane with a location determined by the path's delay and Doppler.
During the stationary time of the doubly-selective channel when this representation is accurate\cite{wcoptvc},
the received OTFS signal in the DD domain can be regarded as the result of a two-dimensional (2D) convolution between the information-bearing symbols and an {effective} {DD domain} channel, which considers the combined effect of the \emph{ideal} transmit pulse, the DD channel and the \emph{ideal} receive pulse\cite{hadani2018otfs}.
As a result, the OTFS modulation can exploit not only frequency diversity but also time diversity of the doubly-selective channel to achieve reliable communication.

Note that the DD plane is also a TF plane, with \emph{fine grids} corresponding to the delay and Doppler resolutions.
The OTFS modulation attempts to couple the transmitted signal with the channel, which means to match the resolutions of the signal modulation to those of the channel in the DD plane \cite{hadani2018otfs,hadaniyt}.
Since the DD plane is divided with specified time (delay) and frequency (Doppler) resolutions, a modulation performed in accordance with the DD plane's resolutions, namely a DD plane modulation, is naturally a multicarrier (MC) modulation with subcarrier spacing being the Doppler resolution.
Considering the information-bearing symbols separately placed  on the DD plane, it is clear that a  DD plane MC modulation requires a pulse orthogonal with respect to the DD plane's resolutions.
For the sake of conciseness, in the remainder of this paper, we call the pulse orthogonal with respect to the DD (or TF) plane's resolutions as DD (or TF) plane orthogonal pulse. Also, without special notice or explanation, the TF plane refers to the conventional signal plane with \emph{coarse grids} adopted in the OFDM modulation.

A pulse confined in one fine grid of the DD plane is obviously a DD plane orthogonal pulse. However, according to the Heisenberg's uncertainty principle, this particular DD plane orthogonal pulse does not exist, which makes it difficult to design a DD plane MC modulation.
In fact, the OTFS modulation may be considered as a practical workaround for this problem.
The OTFS's DD domain signal is first mapped to a TF domain signal {via the inverse symplectic finite Fourier transform (ISFFT),} and then conveyed by the TF plane rectangular pulse, which is a TF plane orthogonal pulse, without violating the Heisenberg's uncertainty principle\footnote{See the OFDM interpretation of OTFS in \cite{hadani2018otfs}.}. In other words, the OTFS signal is orthogonal with respect to the TF plane's coarse resolutions, and its ideal pulse is said to satisfy the \emph{TF plane bi-orthogonal robust property}\cite{Hadani17} with respect to the time and frequency translations induced by the doubly-selective channel. Unfortunately, such a TF plane ideal pulse cannot be realized in practice\cite{hadani2018orthogonal}.
{Because the pulse is the very core of modulation schemes\cite{mct,tff}, without its own pulse, it was pointed out in \cite{zemenpimrc2018} that the OTFS is still an OFDM with an ISFFT precoder. To achieve the coupling between the modulated signal and the DD channel, a DD plane orthogonal MC modulation is a natural and better choice, as its modulated signal's resolutions can be matched to the DD channel's resolutions,
such that each received DD domain signal is simply a linear combination of  the transmit DD domain signals.} However, currently there is no  DD plane orthogonal MC modulation, because whether a realizable DD plane orthogonal pulse exists or not, is still unknown.

In this paper, we investigate the DD plane MC modulation, and answer the above fundamental question for the DD plane MC modulation design.
The key contribution of this paper is to design the new orthogonal pulse with respect to the fine DD resolutions, which to the best of our knowledge, has not been found before.
We provide a rigorous proof of the orthogonality of the new pulse. Then,
based on this newly found DD plane orthogonal pulse, we propose a novel orthogonal delay-Doppler division multiplexing (ODDM) modulation.
We show that in contrast to the OTFS, the proposed ODDM has a compact and exact channel input-output relation in the DD domain, and it can achieve perfect coupling between the modulated signal and the DD channel. Our contributions can be summarized as follows:

\begin{itemize}

	\item By clarifying that the time resolution of a signal plane is symbol interval,  
	      we indicate that a feasible DD plane MC modulation is essentially a type of staggered multitone (SMT) modulation\cite{mct} and that staggering MC symbols provides an opportunity to find a realizable DD plane orthogonal pulse.

	\item Analogous to the OFDM, we propose an ODDM modulation and present it as a staggered upsampled-OFDM in the digital domain. Based on a spectrum analysis of the staggered upsampled-OFDM, we propose a sample-wise square-root Nyquist pulse shaping for the ODDM. We show that {each} ODDM {symbol} can be treated as a pulse-shaped OFDM (PS-OFDM) {symbol}, and we  identify the corresponding transmit pulse.

	\item We prove the \emph{orthogonality} of this particular transmit pulse with respect to the DD plane's resolutions, and show that this pulse is exactly the desired orthogonal pulse for DD plane MC modulation. We reveal that with this transmit pulse and its matched filter as receive pulse, the effective DD domain channel for the ODDM can be \emph{solely and exactly} determined by the {DD} channel.

	\item The associated DD domain channel input-output relation of the proposed ODDM modulation is derived by directly exploiting the well-known frequency domain properties of OFDM symbol with timing and frequency offsets. We show that the  channel matrix has an elegant block-circulant-like structure, which exactly describes the DD domain channel input-output relation of the ODDM.

	\item The performance of the proposed ODDM modulation is demonstrated by simulations. Its superiority over the OTFS is confirmed from performance comparisons in terms of out-of-band emission (OOBE) and bit error rate (BER).

\end{itemize}

The rest of the paper is organized as follows: Section II reviews the OTFS modulation and demodulation. The ODDM is proposed and its staggered upsampled-OFDM representation in digital domain is given in Section III. Next, following a spectral analysis of the staggered upsampled-OFDM, a pulse shaping method for ODDM modulation is proposed, and the corresponding transmit filter is also identified. The ODDM demodulation including matched filtering and signal detection is presented in Section IV. Simulation results are shown in Section V, and finally Section VI concludes the paper.

Notations: In this paper, uppercase boldface letters are used to represent matrices and lowercase boldface letters are used for column vectors. Superscript $\mathcal T$ denotes the transpose operator. Also,  $[\cdot]_M$ stands for the mod $M$ operator, while $\rect_{T}(t)$ denotes the $T$-length rectangular pulse with unit energy.

\section{OTFS modulation and demodulation}

In this section, we briefly review the OTFS modulation and
demodulation. The OTFS modulation \cite{Hadani17,hadani2018orthogonal,hadani2018otfs} is based on the following grid consideration in the TF and DD signal planes:
\begin{itemize}
	\item TF plane grid $\Pi$ :  $\{\grave nT,\grave m\Delta f\}$ for $\grave n=0,\ldots,N-1$ and $\grave m=0,\ldots M-1$, where $N$ refers to the number of {time slots} or OFDM symbols for an OTFS frame, and $M$ refers to the number of subcarriers for the OFDM symbol in each time slot. In addition, each time slot has the duration of $T$ and $\Delta f=\frac{1}{T}$ is the sub-carrier spacing of the OFDM symbol in the TF plane. Here, $\grave m$ and $\grave n$ denote the $\grave m$-th subcarrier and $\grave n$-th OFDM symbol of the OTFS frame, respectively, in the TF plane.

	\item DD plane grid $\Gamma $ :  $\left\{\frac{m}{M\Delta f}, \frac{n}{NT}\right\}$ for $m=0,\ldots,M-1$ and $n=0,\ldots N-1$, {where $M\Delta f$ is the sampling rate of the OTFS signal, leading to the delay resolution of the signal being $\frac{1}{M\Delta f}$, and $NT$ is the total OTFS frame duration, leading to the Doppler resolution of the signal being $\frac{1}{NT}$.} Here $m$ and $n$ denote the $m$-th delay and $n$-th Doppler, respectively.
\end{itemize}
In OTFS, the transceiver uses the transmit and receive pulses $g_{tx}(t)$ and $g_{rx}(t)$, and the cross-ambiguity function (CAF) between them is
\begin{equation}
	A_{g_{tx},g_{rx}}(t,f)=\int g_{tx}(\tilde t) g_{rx}^{*} (\tilde t-t)e^{-j2\pi f(\tilde t-t)} d\tilde t.
\end{equation}
These two pluses are \emph{bi-orthogonal} with respect to translations by time $T$ and frequency $\Delta f$, namely fulfill the perfect reconstruction (PR) condition \cite{mct} of
\begin{equation}\label{prc}
	A_{g_{tx},g_{rx}}(\grave nT,\grave m\Delta f)= \delta(\grave n)\delta(\grave m).
\end{equation}
In addition, they are said to be \emph{orthogonal} when they are matched filters.

\begin{figure*}
	\centering
	\includegraphics[width=0.9\textwidth]{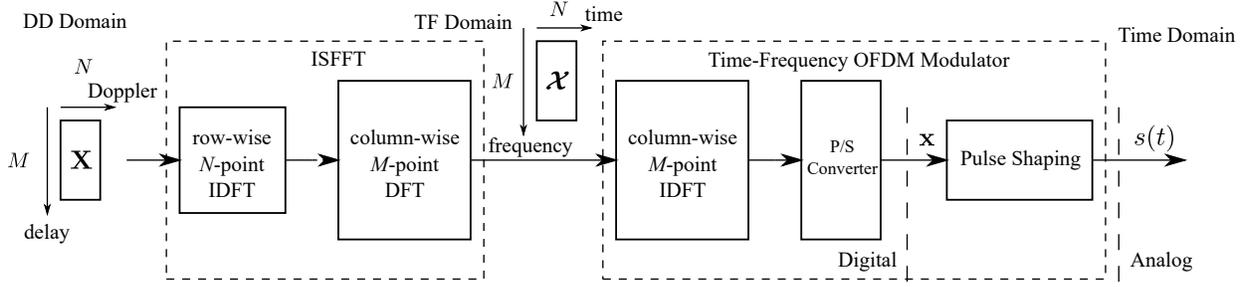}
	\caption{OTFS modulation}
	\label{offgridofdm}
\end{figure*}

The block diagram of OTFS modulation is shown in Fig. \ref{offgridofdm}. Using the ISFFT,
the OTFS transmit symbols $X[m,n]$ in the DD plane are first mapped to the symbols $\mathcal X[\grave m,\grave n]$ in the TF plane as \cite{8424569}
\begin{equation}
	\mathcal X[\grave m,\grave n]=\frac{1}{\sqrt{MN}} \sum_{m=0}^{M-1}\sum_{ n=0}^{N-1} X[m, n] e^{j2\pi\left(\frac{\grave n n}{N}-\frac{\grave m m}{M}\right)}.
\end{equation}
If we arrange the transmitted QAM symbols $X[m, n]$ in an $M\times N$ matrix $\mathbf X$, the ISFFT can be implemented by a row-wise $N$-point inverse discrete Fourier transform (IDFT) and a subsequent column-wise $M$-point DFT. Then, a conventional TF plane OFDM modulator parameterized by the transmit pulse $g_{tx}(t)$ is employed to modulate $\mathcal X[\grave m,\grave n]$ to obtain a continuous time OTFS waveform
\begin{equation}\label{otfswave}
	s(t)=\sum_{\grave n=0}^{N-1}\sum_{\grave m=0}^{M-1} \mathcal X[\grave m,\grave n]g_{tx}(t-\grave nT) e^{j2\pi \grave m\Delta f(t-\grave nT)}.
\end{equation}
The above equation is also known as discrete Heisenberg transform,  or synthesis filter bank in the context of filter bank multicarrier (FBMC) modulation\cite{mct}.

Suppose that a doubly-selective channel is composed of $P$ paths, and it can be written in the DD domain as
\begin{equation}\label{ddchannel}
	h(\tau,\nu)=\sum_{p=1}^P h_p\delta(\tau-\tau_p)\delta(\nu-\nu_p),
\end{equation}
where $\tau$ represents the delay variable, $\nu$ represents the Doppler variable, and $h_p$, $\tau_p$, and $\nu_p$ are the gain, delay, and Doppler of the $p$th path, respectively. {Due to the band-pass filtering and sampling, the observed $h(\tau,\nu)$ is usually modeled by a discrete equivalent channel \cite{tff,6563167} with}
\begin{equation}
	\tau_p =\frac{l_p}{M\Delta f}, \nu_p =\frac{k_p}{NT},
\end{equation}
where $l_p$ and $k_p$ are integers representing the delay and Doppler indices in the DD plane for the $p$-th path, respectively.
Note that this channel model is the classic sampling model for the delay-Doppler spread function with combined time and frequency constraints, which was initially proposed in \cite{bello}.

Since each path introduces a delay and a Doppler to the transmitted waveform $s(t)$, the received signal at time $t$ is given by
\begin{eqnarray}
	r(t)  & = & \sum_{p=1}^P h_p s(t-\tau_p) e^{j2\pi \nu_p(t-\tau_p)} +z(t) \\
	&= & \sum_{p=1}^P\sum_{\grave n=0}^{N-1}\sum_{\grave m=0}^{M-1} h_p\mathcal X[\grave m,\grave n] g_{tx}(t-\grave nT-\tau_p) \nonumber \\
	&& \times e^{j2\pi \left(\grave m\Delta f(t-\grave nT-\tau_p) +\nu_p(t-\tau_p)\right)}  +  z(t),
\end{eqnarray}
where $z(t)$ is the additive white Gaussian noise (AWGN) at time $t$.
At the output of the receive filter, $\mathcal Y[\check m,\check n]$ is obtained by sampling the CAF between the received signal and the receive pulse given by
\begin{equation}\label{caf}
	A_{r,g_{rx}}(t,f)=\int r(\tilde t) g_{rx}^{*} (\tilde t-t)e^{-j2\pi f(\tilde t-t)} d\tilde t
\end{equation}
at the TF grid $\Pi$, namely $\mathcal Y[\check m,\check n]=A_{r,g_{rx}}(\check nT,\check m\Delta f)$, which is also known as discrete Wigner transform, or analysis filter bank in the context of FBMC modulation\cite{mct}.

The TF domain channel input-output relation of the OTFS modulation between the TF domain signals $X[\tilde m, \tilde n]$ and $\mathcal Y[\check m,\check n]$ can be written as \cite{8424569}
\begin{equation}\label{tfh}
	\mathcal Y[\check m,\check n]=\sum_{\tilde m=0}^{M-1}\sum_{\tilde n=0}^{N-1} \mathcal H_{\check m,\check n}[\tilde m, \tilde n] \mathcal X[\tilde m, \tilde n] +\mathcal Z[\check m,\check n],
\end{equation}
where the TF domain channel, denoted by $\mathcal H_{\check m,\check n}[\tilde m, \tilde n]$, is given by
\begin{equation*}
	\begin{split}
		\mathcal H_{\check m,\check n}[\tilde m, \tilde n]  = & \sum_{p=1}^P \{h_p A_{g_{tx},g_{rx}}((\check n-\tilde n)T-\tau_p, (\check m-\tilde m)\Delta f-\nu_p) \\
		& \times e^{j2\pi (\nu_p  +\tilde m \Delta f)((\check n-\tilde n)T-\tau_p)}e^{j2\pi \nu _p \tilde n T}\}.
	\end{split}
\end{equation*}
It corresponds to the ISI and ICI when $(\check m, \check n) \ne (\tilde m, \tilde n)$, and $\mathcal Z[\check m,\check n]$ is the noise term in the TF domain.

Let $\tau_{\max}$ and $\nu_{\max}$ be the maximum delay and Doppler of the doubly-selective channel, respectively. We have $-\tau_{\max} \le \tau_p \le \tau_{\max}$ and  $-\nu_{\max} \le \nu_p \le \nu_{\max}$.
An ideal transmit and receive pulse pair for OTFS is said to satisfy the \emph{bi-orthogonal robust} condition \cite{Hadani17,hadani2018orthogonal} as
\begin{equation}\label{bior}
	A_{g_{tx},g_{rx}}((\check n-\tilde n)T-\tau_p, (\check m-\tilde m)\Delta f-\nu_p)=\delta (\check n-\tilde n) \delta(\check m-\tilde m)
\end{equation}
for all $\tau_p$ and $\nu_p$, which obviously cannot be realized in practice for a general doubly-selective channel.

From (\ref{tfh}), the DD domain channel input-output relation for the OTFS modulation can be obtained by applying the symplectic finite Fourier transform (SFFT) to $\mathcal Y[\check m,\check n]$. For the $T$-length rectangular pulses $g_{tx}(t)=g_{rx}(t)=\textrm{Rect}_{T}(t)$, the DD domain received signal can be approximated as \cite{8424569} %
\begin{equation}\label{ddy}
	\begin{split}
		Y[m,n]\approx & \sum_{p=1}^P h_p e^{j2\pi \left(\frac{m-l_p}{M}\right)\frac{k_p}{N}} \\
		& \times \alpha_p(m,n)X[[m-l_p]_M,[n-k_p]_N]+\zeta [m,n],
	\end{split}
\end{equation}
where $\zeta [m,n]$ is the DD domain noise sample and
\begin{equation*}
	\alpha_p(m,n)=
	\left\{
	\begin{array}{ll}
		1                                                        & l_p\le m \le M \\
		\frac{N-1}{N} e^{-j2\pi\left(\frac{[n-k_p]_N}{N}\right)} & 0\le m \le l_p
	\end{array}
	\right.
\end{equation*}
Furthermore, the DD domain input-output relation can be rewritten in a matrix form as
\begin{equation}\label{otfsiorelation}
	\boldsymbol y\approx \boldsymbol H \boldsymbol x +\boldsymbol \zeta,
\end{equation}
where $\boldsymbol y$ and $\boldsymbol x$ are the vectorized $MN$ received and transmitted DD domain signals, respectively.
Since there are only $P$ nonzero entries in each row and each column of $\boldsymbol H$, a message passing detection algorithm can be employed to recover the information-bearing QAM symbols\cite{8424569}.

From (\ref{prc}) and (\ref{otfswave}), it is clear that although the OTFS considers the signal in the DD plane, it is essentially a TF plane MC modulation based on the TF plane orthogonal pulse.

\begin{figure*}
	\centering
	\includegraphics[width=0.9\textwidth]{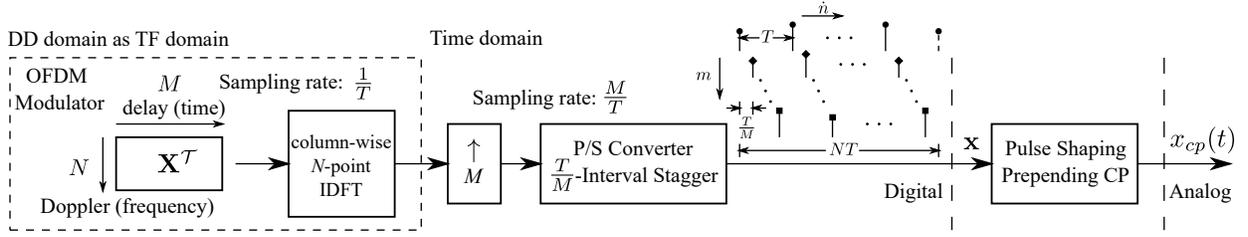}
	\caption{ODDM modulation}
	\label{ongridofdm}
\end{figure*}

\section{ODDM modulation}
Now we introduce the basic concept of ODDM modulation, its modulated digital sequence, and subsequently the ODDM waveform.
Bearing in mind that $\Delta f= \frac{1}{T}$, we can rewrite the TF and DD grids as $\Pi=\{\grave nT,\grave  m\frac{1}{T}\}$ and $\Gamma=\left\{ m\frac{T}{M}, n\frac{1}{NT}\right\}$, respectively. For the sake of comparison, we will replace $\Delta f$ by $\frac{1}{T}$ hereafter. One can see now that the DD plane is also a TF plane however with \emph{fine} grids corresponding to the delay and Doppler resolutions.
To match the resolutions of the signal plane to those of the channel plane\footnote{{Here, we refer to the equivalent DD channel that we can observe.}}, a modulation  performed along with the DD plane's resolutions is preferred.

The fine frequency resolution of the DD plane indicates that we actually need an MC modulation and a DD plane orthogonal pulse. For MC modulation, a common sense so far is that the area of grid (a.k.a. rectangular lattice \cite{mct}) is greater than or equal to $1$, for example, we have $T\times \frac{1}{T}=1$ for the TF grid $\Pi$ corresponding to a realizable transmit pulse that does not violate the Heisenberg uncertainty principle. On the other hand, we have $\frac{T}{M}\times \frac{1}{NT}=\frac{1}{MN} \ll 1$ for the DD grid $\Gamma$, where the obvious DD plane orthogonal pulse confined to $\Gamma$ violates the Heisenberg uncertainty principle and therefore cannot be realized. As a result, it seems questionable to achieve an MC modulation on the DD plane. The solution to this difficulty in the OTFS modulation is to first map the signals from DD plane to TF plane via the ISFFT, then modulate them using the conventional TF plane OFDM, as we can see in (\ref{otfswave}). However, the TF plane rectangular pulse does not satisfy the bi-orthogonal robust condition, while a practical pulse well-localized in the TF plane will cause performance degradation\cite{8516353}. Obviously, a better approach is to design a DD plane MC modulation without mapping the signal to the TF plane.

In the following, we consider the DD plane MC modulation problem. In an MC modulation, each MC symbol is one cycle of a periodic signal, where the symbol period is the inverse of the subcarrier spacing, namely the frequency resolution of signal plane. Hence, the orthogonality among subcarriers can be achieved.  In the absence of CP, the symbol duration is just the symbol period. Meanwhile, the time resolution of signal plane is
symbol interval between consecutive symbols. For the DD plane, its fine grid implies that the symbol period, which is the inverse of the Doppler resolution, is longer than the symbol interval (i.e. the delay resolution). This relation clearly indicates that a feasible DD plane MC modulation is essentially a type of SMT modulation\cite{mct}, where successive MC symbols are staggered (overlapped). In fact, allowing the stagger of MC symbols provides us with an opportunity to find a realizable DD plane orthogonal pulse without violating the Heisenberg's uncertainty principle.

Recall that $\Gamma=\left\{m\frac{T}{M}, n\frac{1}{NT}\right\}$ for $m=0,\ldots,M-1$ and $n=0,\ldots N-1$. For the proposed DD plane MC modulation, we have $MN$ information-bearing symbols to be modulated into $M$ MC symbols, where each MC symbol has $N$ subcarriers. Because the subcarrier spacing is $\frac{1}{NT}$, the symbol period is $NT$. At the same time, these $NT$-length MC symbols are spaced by a short interval $\frac{T}{M}$ corresponding to the delay resolution, which results in a staggered signal structure. Then, the CP-free waveform of the DD plane MC modulation is given by
\begin{equation}\label{ddmcwave}
	x(t)=\sum_{m=0}^{M-1}\sum_{n=0}^{N-1} X[m,n]{\check g}_{tx}\left(t-m\frac{T}{M}\right) e^{j2\pi \frac{n}{NT}\left(t-m\frac{T}{M}\right)},
\end{equation}
where ${\check g}_{tx}(t)$ is the transmit pulse. Clearly, ${\check g}_{tx}(t)$ should be a DD plane orthogonal pulse, that is orthogonal with respect to the DD resolutions $\frac{T}{M}$ and $\frac{1}{NT}$, to avoid the ISI and ICI in the transmit signal $x(t)$. To realize the DD plane MC modulation, a key question now is whether we can find such a realizable DD plane orthogonal pulse ${\check g}_{tx}(t)$, which does not violate the Heisenberg uncertainty principle.

For the time being, let us first assume such a DD plane orthogonal pulse ${\check g_{tx}}(t)$ exists\footnote{We will later {prove} that such a pulse exists.}. With this assumption,
we call the proposed DD plane MC modulation in  (\ref{ddmcwave})   an ODDM modulation.
The ODDM modulation includes $M$ ODDM symbols spaced by $\frac{T}{M}$ but orthogonal with each other; and each ODDM symbol has $N$ orthogonal subcarriers and therefore is effectively an $N$-subcarrier OFDM.
In particular, the symbol interval $\frac{T}{M}$ between the ODDM symbols is much shorter than the symbol period $NT$. Since $M$ ODDM symbols are staggered with interval $\frac{T}{M}$, the ODDM signal has a bandwidth around $\frac{M}{T}$. As we will see later, it is this staggering operation that makes the ODDM different from the conventional OFDM, where the symbol interval is just the symbol period $NT$ and the bandwidth is around $\frac{1}{T}$.

\subsection{ODDM digital sequence}
Before we present the ODDM digital sequence, we briefly discuss the difference between the ODDM and the existing SMT modulations.
A well-known SMT modulation in the literature is the FBMC with offset QAM (OQAM), a.k.a OFDM/OQAM modulation\cite{fbmcprimer,mct}. For comparison, let us consider an $N$-subcarrier OFDM/OQAM modulation with the same subcarrier spacing $\frac{1}{NT}$, where MC symbols occupied the band of $\left(-\frac{1}{2T}, \frac{1}{2T}\right)$ are staggered with an interval of $\frac{NT}{2}$, to compensate the rate loss caused by the real-domain-only orthogonality of the prototype pulse/filter. Observing from the digital domain, we can find that in OFDM/OQAM, the MC symbols are sampled at a rate of $\frac{1}{T}$, and then staggered at an interval of $\frac{NT}{2}$. On the other hand, ODDM samples have a rate of $\frac{1}{T}$, but staggered at an interval of $\frac{T}{M}$. 

Now, let us consider the ODDM digital sequence before being pulse shaped by $\check g_{tx}(t)$. Without loss of generality, for the $m$-th ODDM symbol in (\ref{ddmcwave}), just like the conventional OFDM,  we can use the IDFT to obtain the time-domain discrete samples of the $m$-th ODDM symbol as
\begin{align}\label{moddmtds}
	x[m,\dot n]=\sum_{n=0}^{N-1} X[m,n]e^{j2\pi \frac{\dot n n}{N}}, \dot n=0,\ldots N-1,
\end{align}
where $\dot n$ represents the index of the time-domain discrete samples spaced by $T$.
To stagger $M$ ODDM symbols at an interval of $\frac{T}{M}$, the aforementioned $N$ time-domain discrete samples need to be upsampled by $M$ to obtain $MN$ discrete samples, given by
\begin{align}\label{oddmups}
	\mathbf x[m]= [\overbrace{0,\ldots,0}^{m},
	x[m,0],\overbrace{0,\ldots,0}^{M-1}, x[m,1],\overbrace{0,\ldots,0}^{M-1}, \nonumber \\ \ldots, \overbrace{0,\ldots,0}^{M-1}, x[m,N-1],\overbrace{0,\ldots,0}^{M-m-1}],
\end{align}
which are spaced by $\frac{T}{M}$ and therefore imply that an \emph{analog} ODDM symbol occupies a wideband of $\left(-\frac{M}{2T}, \frac{M}{2T}\right)$ rather than $\left(-\frac{1}{2T}, \frac{1}{2T}\right)$.
By doing this, an ODDM frame consisting of $M$ ODDM symbols only spans a duration around $NT$. Consequently, the ODDM in digital domain can be represented as a staggered upsampled-OFDM, where the upsampling factor and the staggering interval are $M$ and $\frac{T}{M}$, respectively. A block diagram of the ODDM modulation is shown in Fig. \ref{ongridofdm}. \\

\subsection{ODDM waveform design}
It is known that by considering both the ISFFT and the OFDM modulator in Fig. \ref{offgridofdm}, the discrete OTFS samples can be obtained directly using the row-wise $N$-point IDFT\cite{hadaniyt}. Comparing the OTFS in Fig. \ref{offgridofdm} and the ODDM in Fig. \ref{ongridofdm}, it is interesting to find that both of them have the same time domain sample sequence $\mathbf x$ with $MN$ samples at a rate of $\frac{M}{T}$.
However, the physical meanings of this digital sequence has a vital impact on the subsequent pulse shaping and waveform designs, which leads to the fundamental difference between the ODDM and the OTFS waveforms.
In fact, a digital sequence without appropriate pulse shaping is \emph{not} a complete modulation waveform for practical systems. \\

\begin{figure}
	\centering
	\includegraphics[width=8cm]{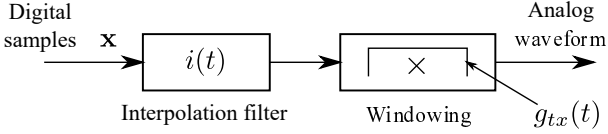}
	\caption{Pulse shaping of OFDM modulation}
	\label{ps_ofdm}
\end{figure}

\subsubsection{Pulse shaping consideration}
For the OTFS, we treat $\mathbf x$ as $MN$ samples of $N$ consecutive $M$-subcarrier OFDM symbols. As shown in Fig. \ref{ps_ofdm}, the pulse shaping is performed per OFDM symbol (per $M$ samples) by a digital-to-analog converter (DAC) and a subsequent windowing function, parameterized by an interpolation filter $i(t)$ and the transmit pulse $g_{tx}(t)$, respectively.
Consider that each OTFS signal has $N$ time slots
corresponding to the $N$ OFDM symbols. The OTFS pulse shaping needs to apply $N$ rectangular pulses to the transmitted signal in $N$ time slots, similar to conventional OFDM signals.
Therefore, the pulse shaping in OTFS is \emph{slot-wise}.
Given that the ideal interpolation filter is the Sinc function $i(t)=\sinc\left(\frac{Mt}{T}\right)$, the transmit pulse $g_{tx}(t)$ is the $T$-length rectangular pulse $\textrm{Rect}_{T}(t)$, which however will cause significant discontinuities among OFDM symbols and consequently serious OOBE\cite{mct}. As an MC modulation, OFDM modulation is also based on the assumption that each symbol is one cycle of a periodic signal. Hence, a practical solution is to add a CP and a cyclic suffix to each symbol, where the length of CP is set longer to contain the dispersive channel and avoid the ISI. Then, a corresponding wider but smoother window function is employed to shape the transmit waveform and suppress the OOBE. Obviously,  introducing CP and cyclic suffix per OFDM symbol (slot) greatly deteriorates the spectrum efficiency.

For the ODDM, the pulse shaping procedure is basically based on that of the OFDM. However, since we treat $\mathbf x$ as $MN$ samples of $M$ staggered $N$-subcarrier OFDM symbols, each symbol has the same length $NT$ as the whole frame, which clearly indicates that the symbol-wise pulse shaping is exactly a \emph{frame-wise} pulse shaping. Furthermore, because the $M$ symbols are staggered, the discontinuity among symbols is not an issue anymore. As a result, the symbol-wise CP is not necessary. To understand the rationale behind this finding, we need to analyze the spectrum of the staggered upsampled-OFDM considering the effect of the DAC's interpolation.

\subsubsection{Spectrum analysis}
For a conventional $N$-subcarrier OFDM symbol with a subcarrier spacing of $\frac{1}{NT}$, we usually pass its $T$-spaced time domain discrete samples through the ideal interpolation filter $i(t)=\sinc\left(\frac{t}{T}\right)$ and then apply the $NT$-length rectangular windowing, namely the rectangular pulse $\textrm{Rect}_{NT}(t)$, to obtain a narrow-band analog OFDM symbol \emph{roughly} banded to $\left(-\frac{1}{2T}, \frac{1}{2T}\right)$.
The unusual point in the ODDM is that in order to stagger {$M$ OFDM symbols}, we do upsample these $T$-spaced  discrete samples by $M$, as shown in (\ref{oddmups}). Thus, the bandwidth is increased by  $M$ times, and we need to pass the corresponding  $\frac{T}{M}$-spaced discreet samples through the ideal interpolation filter $\sinc\left(\frac{Mt}{T}\right)$.

\begin{figure}
	\centering
	\includegraphics[width=8.8cm]{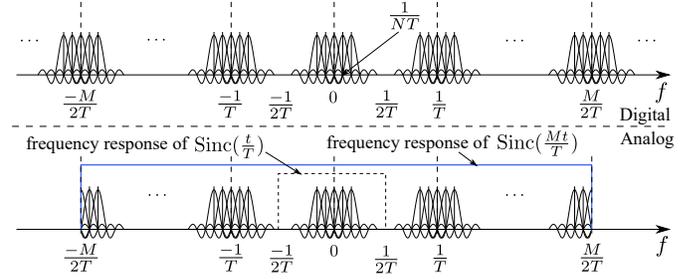}
	\caption{Spectrum of a single ODDM symbol before and after the pulse shaping}
	\label{su-oddm-spec}
\end{figure}

For each ODDM symbol, the time domain $N$ discrete samples given in (\ref{moddmtds}) {are identical to those of a conventional OFDM symbol, which can be obtained by a $\rect_{NT}(t)$-based pulse shaping followed by a rate $\frac{1}{T}$ sampler.} As shown in the upper part of Fig. \ref{su-oddm-spec}, the aliasing caused by the rate $\frac{1}{T}$ sampling spreads the spectrum of the conventional $N$-subcarrier analog OFDM symbol 
over the frequency axis, where several edge subcarriers are deliberately unplotted and the $\sinc$ shape of the rectangular pulse's spectrum  for each subcarrier is truncated for display purpose. Then, after the upsampling to obtain (\ref{oddmups}) and then the convolution with the ideal interpolation filter $\sinc\left(\frac{Mt}{T}\right)$, the spectrum is limited to $\left(-\frac{M}{2T}, \frac{M}{2T}\right)$, as shown in the lower part of Fig. \ref{su-oddm-spec}, where the conventional OFDM signal spectrum limited to $\left(-\frac{1}{2T}, \frac{1}{2T}\right)$ is also pointed out for comparison.

\subsubsection{ODDM waveform}

Note that each ODDM symbol has a similar spectrum representation to that in Fig. \ref{su-oddm-spec}. Staggering them in time domain does not change the overall spectrum occupancy. Therefore, the spectrum of the ODDM waveform depends on the interpolation filter, and the pulse shaping of the ODDM is simplified to the selection of the interpolation filter.

The modulated message is carried inside the narrow band of $\left(-\frac{1}{2T}, \frac{1}{2T}\right)$. In principle, we can choose any interpolation filter with a flat frequency response over $\left(-\frac{1}{2T}, \frac{1}{2T}\right)$. Meanwhile, since $M$ ODDM symbols staggered by $\frac{T}{M}$ between arbitrary two adjacent symbols are completely overlapped in the frequency domain, to decouple them in the time domain, the interpolation filter needs to be ISI-free for the symbol interval of $\frac{T}{M}$. In summary, the pulse shaping of ODDM modulation may be achieved by choosing a suitable interpolation filter, which has
\begin{itemize}
	\item flat frequency response over $\left(-\frac{1}{2T}, \frac{1}{2T}\right)$,
	\item ISI-free impulse response for the symbol interval of $\frac{T}{M}$.
\end{itemize}
As an ISI-free impulse response for the symbol interval of $\frac{T}{M}$ corresponds to a flat frequency response over a wide frequency range close to $\left(-\frac{M}{2T}, \frac{M}{2T}\right)$, we can simply employ a Nyquist pulse for the symbol interval of $\frac{T}{M}$, which indicates that the pulse shaping of ODDM modulation can be performed in a \emph{sample-wise} manner. At the same time, it is interesting to notice that the Nyquist pulse is exactly what we use in conventional single carrier communication, which inspires us that it can be divided in the frequency domain, one for the transmit pulse and one for the receive pulse to perform the matched filtering.

From the above analysis, we propose a pulse shaping method for ODDM modulation, where the pulse shaping is performed sample-wise by a square-root Nyquist pulse for the symbol interval of $\frac{T}{M}$.
For such a Nyquist pulse, there are many candidates in the literature, including the well-known raised cosine pulses. Without loss of generality, suppose that a time-symmetric real-valued square-root Nyquist pulse $a(t)$ is employed as the interpolation filter $i(t)$, where $\int_{-\infty}^{+\infty}|a(t)|^2 dt =\frac{1}{N}$. 

Let $a(t)$ span a time duration of $2Q\frac{T}{M}$, where $Q$ is an integer and $2Q \ll M$, and $a(t)=0$ for $t\notin \left(-Q\frac{T}{M} , Q\frac{T}{M}\right)$. The time domain signal for the $m$-th ODDM symbol generated using the proposed sample-wise pulse shaping becomes
\begin{equation}\label{xm}
	x_m(t)=\sum_{\dot n=0}^{N-1} \sum_{n=0}^{N-1} X[m,n] e^{j2\pi \frac{\dot n n}{N}} a(t-\dot nT),
\end{equation}
where $\dot n$ and $n$ are the indices of time and frequency (Doppler), respectively.
Then, the whole ODDM frame, which consists of $M$ ODDM symbols and spans over the time interval of $-Q\frac{T}{M} \le t \le NT+(Q-1)\frac{T}{M}$, is represented by
\begin{eqnarray}\label{xt}
	x(t) & = & \sum_{m=0}^{M-1} x_m\left(t-m\frac{T}{M}\right) \\
	& = & \sum_{m=0}^{M-1}\sum_{\dot n=0}^{N-1} \sum_{n=0}^{N-1} X[m,n] e^{j2\pi\frac{ \dot n n}{N}} a\left(t-m\frac{T}{M}-\dot nT\right). \nonumber
\end{eqnarray}

\subsubsection{Transmit pulse of ODDM}
The ODDM signal in (\ref{xt}) is generated by filtering the discrete staggered upsampled-OFDM signals with a square-root Nyquist filter, which is a simple way to generate the analog ODDM waveform. However, this method is different from the conventional way to generate OFDM or PS-OFDM waveform, as shown in Fig. \ref{ps_ofdm}. Obviously, the ODDM can be considered as a PS-OFDM\cite{mct}. With the proposed sample-wise square-root Nyquist pulse shaping, a \emph{fundamental} question is that from the view point of PS-OFDM, what is the equivalent transmit pulse $\check g_{tx}(t)$ for the ODDM now?

\begin{figure}
	\centering
	\includegraphics[width=8.8cm]{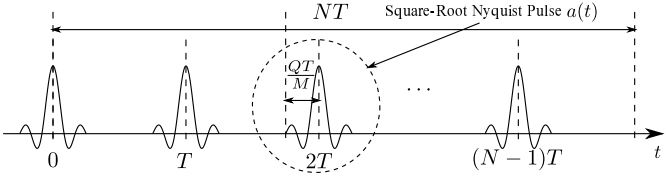}
	\caption{Transmit pulse $u(t)$}
	\label{ps-su-odfdm}
\end{figure}

As shown in Fig. \ref{ps-su-odfdm}, we can form a transmit pulse $u(t)$ as
\begin{equation}\label{ut}
	u(t)=\sum_{\dot n=0}^{N-1}a(t-\dot nT).
\end{equation}
{Because of $\int_{-\infty}^{+\infty}|a(t)|^2 dt =\frac{1}{N}$, we have $\int_{-\infty}^{+\infty}|u(t)|^2 dt=1$.}
In the context of PS-OFDM, the OFDM waveform with the transmit pulse $u(t)$ is
\begin{eqnarray}
	\tilde x_m(t) & = & \sum_{n=0}^{N-1} X[m,n] e^{j2\pi\frac{ n t}{NT}}  u(t), \nonumber \\
	& = & \sum_{n=0}^{N-1} X[m,n] e^{j2\pi\frac{ n t}{NT}} \times \sum_{\dot n=0}^{N-1}a(t-\dot nT). \label{dotxm}
\end{eqnarray}

Now we have the following lemma.

\begin{lemma}
	With the proposed sample-wise square-root Nyquist pulse shaping using $a(t)$, the generated ODDM symbol is a {near-perfect} approximation of a PS-OFDM symbol, whose transmit pulse is $u(t)$.
\end{lemma}
\begin{IEEEproof}
	See Appendix A.
\end{IEEEproof}

\begin{figure}
	\centering
	\includegraphics[width=8.8cm]{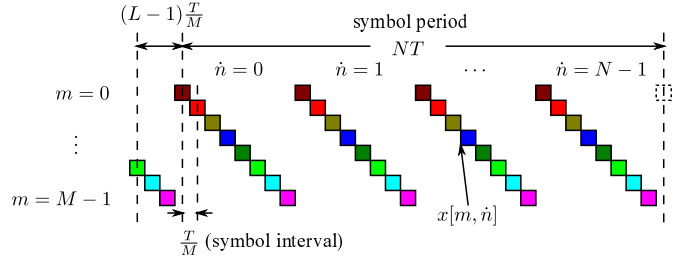}
	\caption{ODDM frame structure, $M=8$, $N=4$, $L=4$.}
	\label{oddm-frame}
\end{figure}

From Lemma 1, we know that after the proposed sample-wise square-root Nyquist pulse shaping using $a(t)$, each ODDM symbol can be approximately considered as a PS-OFDM symbol being pulse-shaped by $u(t)$. In other words, the equivalent transmit pulse is $u(t)$, compared to the classic $NT$-length rectangular pulse $\rect_{NT}(t)$ which will generate a narrow-band OFDM symbol roughly banded to $\left(-\frac{1}{2T}, \frac{1}{2T}\right)$.
We then have $M$ staggered $N$-subcarrier PS-OFDM symbols pulse-shaped by $u(t)$ to form an ODDM frame
\begin{eqnarray}\label{xtut}
	x(t) = \sum_{m=0}^{M-1}\sum_{n=0}^{N-1} X[m,n] u\left(t-m\frac{T}{M}\right) e^{j2\pi\frac{ n }{NT}\left(t-m\frac{T}{M}\right)},
\end{eqnarray}
where $-Q\frac{T}{M} \le t \le NT+(Q-1)\frac{T}{M}$. Furthermore, considering the channel delay spread,
an $(L-1)\frac{T}{M}$-length of CP is prepended to the head of the frame. An example of the proposed time domain frame structure for $M=8$, $N=4$, $L=4$ is shown in Fig. \ref{oddm-frame}, where each block means one time domain sample $x[m,\dot n]$ given in (\ref{moddmtds})
and the blocks with the same color form one ODDM symbol. Finally, considering the prepended CP, we can extend the definition of $u(t)$ to
\begin{equation}\label{ucpt}
	u_{cp}(t)=\sum_{\dot n=-1}^{N-1}a(t-\dot nT).
\end{equation}
Then, the CP-included ODDM waveform spanning over $-(L+Q-1)\frac{T}{M} \le t \le NT+(Q-1)\frac{T}{M}$ becomes
\begin{eqnarray}\label{xtcp}
	x_{cp}(t) = \sum_{m=0}^{M-1}\sum_{n=0}^{N-1} X[m,n] u_{cp}\left(t-m\frac{T}{M}\right) e^{j2\pi\frac{ n }{NT}(t-m\frac{T}{M})},
\end{eqnarray}
where $u_{cp}(t)=u(t)$ for $t\in \left(-Q\frac{T}{M}, (N-1)T+Q\frac{T}{M}\right)$ and $u_{cp}(t)=0$ for $t\in \left(-T+Q\frac{T}{M}, -Q\frac{T}{M}\right)$.

\begin{remark}
	From (\ref{xtut}), one can see that the ODDM has a standard MC modulation signal format: The DD plane signal $X[m,n]$ is directly conveyed by the transmit pulse $u(t)$, while the symbol interval and the subcarrier spacing are $\frac{T}{M}$ and $\frac{1}{NT}$, respectively.
	On the other hand, for OTFS modulation in (\ref{otfswave}), the DD plane signal $X(m,n)$ is first mapped to the TF plane signal $\mathcal X[\grave m,\grave n]$ via the ISFFT. Then, the TF plane signal $\mathcal X[\grave m,\grave n]$ is conveyed by the transmit pulse $g_{tx}(t)=\rect_{T}(t)$, while the symbol interval and the subcarrier spacing are $T$ and $\frac{1}{T}$, respectively. Although the ODDM and the OTFS have the same discrete representation for the time interval of $\frac{T}{M}$, it becomes clear that fundamentally they are two different modulation schemes with different waveforms in (\ref{xtut}) and (\ref{otfswave}), respectively.
\end{remark}

\begin{remark}
	Each ODDM symbol can also be approximately considered as a filtered OFDM symbol, where the filter is a \emph{wideband} square-root Nyquist filter with a passband around $\left(-\frac{M}{2T}, \frac{M}{2T}\right)$.
	By employing the wideband square-root Nyquist filter, the ODDM modulation essentially spreads $N$ information-bearing subcarriers within $\left(-\frac{1}{2T}, \frac{1}{2T}\right)$  $M$ times over the overall bandwidth, and therefore has a potential maximum frequency diversity gain of $M$. Meanwhile, since each symbol has a long period of $NT$, the ODDM modulation can also have a potential maximum time diversity gain of $N$. Thus, the ODDM can explore both time and frequency diversities. Note that the achievable diversity of the proposed ODDM is determined by the number of independent paths of the channel, just like that of OTFS in \cite{2010.03344}.
\end{remark}

\section{ODDM Demodulation}
After receiving an ODDM frame distorted by the doubly-selective channel, the receiver will perform the matched filtering to recover the transmitted signal. Until now, the ODDM waveform is presented without the proof of the orthogonality. Here, we show that $u(t)$ is orthogonal with respect to the DD plane's resolutions.
\begin{lemma}\label{l1}
	$u(t)$ satisfies the orthogonal property that
		{
			\begin{equation}
				A_{u,u}\left(m\frac{T}{M}, n\frac{1}{NT}\right)= \delta(m)\delta(n),
				\label{orth}
			\end{equation}}
	for $|m|\le M-1$ and $|n| \le N-1$.
\end{lemma}
\begin{IEEEproof}
	See Appendix B.
\end{IEEEproof}

From Lemma 2 and by comparing (\ref{ddmcwave}) to (\ref{xtut}), one can see that $u(t)$ is exactly the DD plane orthogonal pulse $\check g_{tx}(t)$ we are looking for.
From the proof, one can see that as long as $2Q\ll M$, any pulse can satisfy the orthogonality with respect to the Doppler resolution of $\frac{1}{NT}$.
Meanwhile, it is the ISI-free property of the Nyquist filter that guarantees the orthogonality with respect to the delay resolution of $\frac{T}{M}$. {Now it becomes clear that because the DD plane orthogonal pulse needs to be Nyquist with interval $\frac{T}{M}$, and at the same time has a period of $NT$, it must be \emph{locally wideband and globally narrow-band}.
		The pulse train $u(t)$ is composed of $N$ equally distributed square-root Nyquist pulses, and it perfectly meets the aforementioned requirements, where a single pulse $a(t)$ is a wideband pulse, while the total $N$ pulses spanning a long period of $NT$ can be virtually treated as a narrow-band pulse.} In addition, similar to what we have done in the sample-wise pulse shaping, the matched filtering using $u(t)$ can also be implemented approximately by a sample-wise matched filtering by $a(t)$ followed by an $N$-point DFT.

From (\ref{orth}) of Lemma 2, it is obvious that $u(t)$ fulfills the PR condition with respect to the fine grid $\Gamma$ of the DD plane. It is noteworthy that this orthogonality is different from the bi-orthogonality in (\ref{prc}) with respect to the coarse resolutions $\Pi$ of the TF plane.
This result is very important and it makes the proposed ODDM differ from the OTFS. Recall that in the DD plane which is used to represent the DD channel impulse response, each channel path introduces delay and Doppler with the resolutions of $\frac{T}{M}$ and $\frac{1}{NT}$, respectively. Because the $T$-length rectangular pulse of the TF plane OFDM fulfills the PR condition with respect to the grid $\Pi$, under the channel's delay and Doppler which are \emph{fractional} to $\Pi$, the bi-orthogonal robust condition in (\ref{bior}) cannot be satisfied in practice. As a result, these fractional channel distortions corresponding to $\Pi$ creates \emph{blurred} ISI and ICI in the TF domain and subsequently complicate the interference in the DD domain, see approximations in the derivation of OTFS DD channel input-output relation (\ref{ddy}) \cite{8424569}.
In contrast, the pulse $u(t)$ of the ODDM
fulfills the PR condition with respect to the fine grid $\Gamma$. Since the DD channel taps' delay and Doppler are \emph{integer multiples} of delay resolution and Doppler resolution, the effective DD domain channel, which considers the combined effect of the transmit pulse $u(t)$, the DD channel  and the receive pulse $u(t)$, can be solely and exactly represented by the DD channel.  The deviation of the DD domain channel matrix can be significantly simplified and exactly determined, as we will see in the associated DD domain channel input-output relation in the following.

\subsection{Input-Output channel relation in DD domain}
From (\ref{xtcp}), we have the received ODDM signal for $-(L+Q-1)\frac{T}{M} \le t \le NT+(L+Q-2)\frac{T}{M}$ as
\begin{eqnarray}
	y(t) & = & \sum_{p=1}^{P} h_p x_{cp}(t-\tau_p) e^{j2\pi \nu_p (t-\tau_p)}+z(t), \nonumber  \\
	& = & \sum_{p=1}^{P} \sum_{m=0}^{M-1}\sum_{n=0}^{N-1} h_p X[m,n] u_{cp}\left(t-m\frac{T}{M}-\tau_p\right)  \nonumber \\
	&& \times e^{j2\pi \left(\frac{n}{NT}(t-m\frac{T}{M}-\tau_p)+\nu_p(t-\tau_p)\right)}+z(t), \nonumber \\
	& = & \sum_{p=1}^{P} \sum_{m=0}^{M-1}\sum_{n=0}^{N-1} h_p X[m,n] u_{cp}\left(t-(m+l_p)\frac{T}{M}\right)  \nonumber \\
	&& \times e^{j2\pi \left(\frac{n}{NT}(t-(m+l_p)\frac{T}{M})+\frac{k_p}{NT}(t-l_p\frac{T}{M})\right)}+z(t) \nonumber, \\
	& = & \sum_{p=1}^{P} \sum_{m=0}^{M-1}\sum_{n=0}^{N-1} h_p X[m,n] u_{cp}\left(t-(m+l_p)\frac{T}{M}\right)  \nonumber \\
	&& \times e^{j2\pi \frac{(n+k_p)}{NT}\left(t-(m+l_p)\frac{T}{M}\right)} e^{j2\pi\frac{k_p m}{MN}}+z(t).
\end{eqnarray}
Owing to the orthogonality of $u(t)$ and $u_{cp}(t)=u(t)$ for $t\in \left(-Q\frac{T}{M}, (N-1)T+Q\frac{T}{M}\right)$, after the matched filtering using $u(t)$, we obtain the signal at the $n$-th subcarrier of the $m$-th ODDM symbol as
\begin{eqnarray}\label{Ymn0}
	Y(m,n) & =  & \int y(t) u\left(t-m\frac{T}{M}\right) e^{-j2\pi\frac{ n }{NT}\left(t-m\frac{T}{M}\right)} dt, \nonumber \\
	& = & \sum_{p=1}^{P}  h_p \tilde X[\hat m,\hat n] e^{j2\pi\frac{k_p (m-l_p)}{MN}}+z[m,n], \label{DDsignal}
\end{eqnarray}
where $\tilde X[\hat m,\hat n] = X[\hat m,\hat n]$ for $\hat n =[n-k_p]_N$ and $\hat m = m-l_p \ge 0$, and $z[m,n]$ is the DD domain noise sample.
When $\hat m = m-l_p < 0$, because of the CP,
the $\hat m$-th symbol is just a $T$ cyclic time-shift of the $(M+\hat m)$-th symbol.  Since $T$ cyclic time-shift of a PS-OFDM symbol with a subcarrier spacing of $\frac{1}{NT}$ corresponds to a phase rotation term $e^{-j2\pi\frac{\hat n}{NT}T}=e^{-j2\pi\frac{\hat n}{N}}$ applied to its frequency domain signal, for $\hat m = m-l_p < 0$ in (\ref{DDsignal}), we have
\begin{equation}\label{xmincp}
	\tilde X[\hat m,\hat n] = e^{-j2\pi\frac{\hat n}{N}} X[M+\hat m,\hat n].
\end{equation}
The sample-wise results in (\ref{DDsignal}) can be vectorized to obtain a more insightful symbol-wise DD channel input-output relation.

Without loss of generality, assume that the maximum delay and Doppler of the channel are  $(L-1)\frac{T}{M}$ and $K\frac{1}{NT}$, respectively.
The $P$ paths can be arranged in a $(2K+1)\times L$ DD domain channel matrix $\mathbf G$, where each row and each column of $\mathbf G$ correspond to a Doppler and delay index, respectively. For example, let $\hat k=k-K-1$, a non-zero element of $\mathbf G$, denoted by $g(\hat k+K+1,l)$, equals to the gain $h_p$ of the $p$th path, whose delay and Doppler are $l\frac{T}{M}$ and $\hat k \frac{1}{NT}$, respectively. Clearly, the total number of non-zero elements in $\mathbf G$ is $P$.

From (\ref{xm}), we can use an $N\times 1$ vector $\mathbf x_m=[X[m,0],\ldots, X[m,N-1]]^{\mathcal T}$ to represent the frequency (i.e. Doppler) domain signals of the $m$-th transmitted ODDM symbol, namely the $m$-th transmitted signal vector in the DD domain because $m$ is the index of delay. Bear in mind that the matched filtering using $u(t)$ in (\ref{Ymn0}) can be approximately implemented by a $a(t)$-based sample-wise matched filtering followed by an $N$-point DFT. At the receiver, after the matched filtering based on $a(t)$ and by discarding the frame-wise CP, we have $MN$ digital samples which can be downsampled by a factor of $M$ to obtain $M$ sequences, representing the time domain signals of $M$ received ODDM symbols ($N$ samples per symbol), respectively. Then, we can apply the $N$-point DFT to them to obtain the corresponding frequency (i.e. Doppler) domain signals, denoted by $\mathbf y_m$ for $0 \le m \le M-1$. These $M$ frequency domain signal vectors are the received DD domain signals.

Recall that due to the orthogonality of the transmit pulse, the effective DD domain channel for the ODDM can be solely represented by the DD channel.
As each ODDM symbol is a PS-OFDM symbol, the interference among the ODDM symbols can be obtained directly using the well-known \emph{frequency domain} properties of OFDM symbol with \emph{integer} timing and frequency offsets with respect to symbol interval and subcarrier spacing, respectively.

From Fig. \ref{oddm-frame}, we can observe that for the $m$-th received ODDM symbol, the path with a delay of  $l\frac{T}{M}$ brings an ISI from the $(m-l)$-th ODDM symbol, where the path's Doppler $\hat k\frac{1}{NT}$ cyclically shifts the subcarrier of the interfering $(m-l)$-th ODDM symbol by $\hat k$. Also, since the $(m-l)$-th ODDM symbol starts from $\frac{ (m-l)T}{M}$, the Doppler also introduces a phase rotation $e^{j2\pi\hat k\frac{1}{NT}\frac{ (m-l)T}{M}}=e^{j2\pi\frac{\hat k(m-l)}{MN}}$. As a result, for the $m$-th received ODDM symbol, the ISI from the $(m-l)$-th ODDM symbol, which is introduced by all paths with the same delay of $l\frac{T}{M}$ but different Dopplers of $\hat k\frac{1}{NT}$, can be governed by
\begin{eqnarray}
	\mathbf H_l^m & = & \sum_{\hat k=-K}^{K} g(\hat k+K+1,l) e^{j2\pi \frac{\hat k(m-l)}{MN} } \mathbf C^{\hat k},
\end{eqnarray}
where $\mathbf C$ is the $N \times N$ cyclic permutation matrix
\begin{equation}
	\mathbf C=
	\begin{bmatrix}
		0      & \ldots & 0      & 1      \\
		1      & \ddots & 0      & 0      \\
		\vdots & \ddots & \ddots & \vdots \\
		0      & \ldots & 1      & 0
	\end{bmatrix}.
\end{equation}

Meanwhile, due to the prepended CP, a negative $m-l$ indicates that the ISI is from the $[m-l]_M$-th ODDM symbol.
Also, because the first sample of the interfering $[m-l]_M$-th ODDM symbol is located in the CP,
which indicates that it is cyclically time-shifted by $T$, there is an additional phase rotation $e^{-j\frac{2\pi}{N} n}$ applied to its $n$th subcarrier.

From the above analysis, we know that for each received ODDM symbol $\mathbf y_m$, the signal term from $\mathbf x_{m-l}$ is $\mathbf H_l^m \mathbf x_{m-l}$, for $0\le l \le L-1$. When $m-l<0$, like (\ref{xmincp}), additional phase rotation term $\mathbf D$ is applied to $\mathbf x_{[m-l]_M}$, where
\begin{equation}
	\mathbf D = \textrm{diag}\left\{1, e^{-j\frac{2\pi}{N}},\ldots, e^{-j\frac{2\pi(N-1)}{N}}\right\}.
\end{equation}
Therefore, the input-output relation in the DD domain for the ODDM modulation can be written in a matrix form at the top of next page, 
\begin{figure*}
	\begin{eqnarray}\label{iodd}
		\begin{bmatrix}
			\mathbf y_0 \\
			\mathbf y_1 \\
			\vdots      \\
			\mathbf y_{M-1}
		\end{bmatrix}
		& = &
		\begin{bmatrix}
			\mathbf H_{L-1}^0 & \cdots & \mathbf H_0^0 &                       &        & \mbox{\Huge 0}     \\
			                  & \ddots & \ddots        & \ddots                &        &                    \\
			                  &        & \ddots        & \ddots                & \ddots &                    \\
			\mbox{\Huge 0}    &        &               & \mathbf H_{L-1}^{M-1} & \cdots & \mathbf H_0 ^{M-1}
		\end{bmatrix}
		\begin{bmatrix}
			\mathbf D \mathbf x_{M-L+1} \\
			\vdots                      \\
			\mathbf D \mathbf x_{M-1}   \\
			\mathbf x_0                 \\
			\vdots                      \\
			\mathbf x_{M-L+1}           \\
			\vdots                      \\
			\mathbf x_{M-1}
		\end{bmatrix}
		+
		\begin{bmatrix}
			\mathbf z_0 \\
			\mathbf z_1 \\
			\vdots      \\
			\mathbf z_{M-1}
		\end{bmatrix},
	\end{eqnarray}
\end{figure*}
where $\mathbf z_m$ stands for the noise. We can rewrite (\ref{iodd}) as
\begin{equation} \label{ioddcompact}
	\mathbf y=\mathbf H \mathbf x + \mathbf z,
\end{equation}
where
$\mathbf y  =  [\mathbf y_0^{\mathcal T},\mathbf y_1^{\mathcal T}, \cdots, \mathbf y_{M-1}^{\mathcal T} ]^{\mathcal T}$, $\mathbf x  =  [\mathbf x_0^{\mathcal T},\mathbf x_1^{\mathcal T}, \cdots, \mathbf x_{M-1}^{\mathcal T} ]^{\mathcal T}$,
$\mathbf z  =  [\mathbf z_0^{\mathcal T},\mathbf z_1^{\mathcal T}, \cdots, \mathbf z_{M-1}^{\mathcal T} ]^{\mathcal T}$, and $\mathbf H$ is the equivalent DD domain channel with size $MN \times MN$ given by
\begin{equation}\label{H}
	\mathbf H=
	\begin{bmatrix}
		\mathbf H_0^0         &        &        &                       & \mathbf H_{L-1}^0 \mathbf D & \cdots         & \cdots & \mathbf H_{1}^0  \mathbf D      \\
		\vdots                & \ddots &        &                       &                             & \ddots         & \ddots & \vdots                          \\
		\vdots                & \ddots & \ddots &                       &                             &                & \ddots & \vdots                          \\
		\mathbf H_{L-2}^{L-2} & \ddots & \ddots & \mathbf H_{0}^{L-2}   &                             & \mbox{\Huge 0} &        & \mathbf H_{L-1}^{L-2} \mathbf D \\
		\mathbf H_{L-1}^{L-1} & \ddots & \ddots & \ddots                & \mathbf H_{0}^{L-1}         &                                                           \\
		                      & \ddots & \ddots & \ddots                & \ddots                      & \ddots                                                    \\
		                      &        & \ddots & \ddots                & \ddots                      & \ddots         & \ddots                                   \\
		\mbox{\Huge 0}        &        &        & \mathbf H_{L-1}^{M-1} & \ddots                      & \ddots         & \ddots & \mathbf H_{0}^{M-1}
	\end{bmatrix}.
\end{equation}

{For each $\mathbf H_l^m$, if there is only one path at that delay $l$, $\mathbf H_l^m$ is simply an $N\times N$ cyclic shift permutation matrix up to a scale factor. Therefore, it has only one non-zero element at each row and each column. Moreover, even there are $P_l>1$ paths with different Dopplers but the same delay $l$, $\mathbf H_l^m$ is still an $N\times N$ circulant matrix, where each row and each column has $P_l$ non-zero elements. As a result, each row and each column of $\mathbf H$ has only $P=\sum_{l=0}^{L-1}P_l$ non-zero elements, and therefore $\mathbf H$ is generally a sparse matrix when $MN\gg P$.}
Meanwhile, it can be observed that regardless of the sparsity, the channel matrix $\mathbf H$ in (\ref{H}) has an elegant block-circulant-like structure, which can be exploited in signal detection.\footnote{This will be discussed in our future work.} On the other hand, after the local matched filtering by $a(t)$, the successive noise samples become correlated and therefore $\mathbf z$ is a colored Gaussian noise vector.
However, because of $2Q\ll M$, the elements of $\mathbf z_m$ are sufficiently distant to be irrelevant, $\mathbf z_m$ is still a white Gaussian noise vector.

In the standard CP-OFDM modulation, it is well-known that the time domain channel input-output relation for each symbol can be represented by a circulant channel matrix composed by the channel impulse response, whose rows and columns correspond to time and delay indices, respectively. The frequency domain counterpart is a diagonal matrix, whose main diagonal is the channel frequency response, where both the rows and the columns now correspond to frequency index, because the effect of channel delay has been absorbed by the CP. On the other hand, in the ODDM modulation, the DD domain channel matrix is used to represent the relation between delay and Doppler domains. Therefore, although $\mathbf H$ in (\ref{H}) has a similar block-circulant-like structure, the physical meaning is completely different. Notice that $\mathbf x_m$ and $\mathbf y_m$ are both frequency domain signal vectors, the rows and columns of $\mathbf H$ actually correspond to Doppler (frequency) and delay indices in a block-wise fashion for $M$ symbols. Also, because the ODDM is essentially a type of SMT modulation, the ISI introduced by channel delay cannot be absorbed by the frame-wise CP. We then observe the unavoidable ISI components from the non-zero off-diagonal block-elements of $\mathbf H$, where the Doppler induced ICI is embedded.

\begin{remark}
	In the design of MC systems, to combat the time and frequency dispersion effects of the doubly-selective channel, we usually add redundancy at a price of spectrum efficiency loss, for example, inserting CP and increasing subcarrier spacing. Another approach is to optimize transmit and receive pulses, see \cite{mct,oodtm,wssusp,aoili,tff} and references therein. On the other hand, the OTFS modulation shows us a novel way to handle these two dispersion effects by {considering the discretization of not only delay but also Doppler.} In a sense, one of the most important discoveries of the OTFS may be {the utilization of the discretization of Doppler}, as we are familiar with that of delay in the conventional wireless channel model. As long as the DD channel dispersion is \emph{on-the-grid} with respect to the fine grid $\Gamma$, it is desirable to design an MC modulation whose signal grid is $\Gamma$, which obviously requires a pulse fulfilled the PR condition with respect to $\Gamma$. By doing this, we can avoid annoying fractional time and frequency offsets which bring blurred ISI and ICI. 
		{In fact, with the sparsely distributed square-root Nyquist pulse train $u(t)$, the proposed ODDM is exactly designed under this criterion, where the ISI and ICI caused by the doubly-selective channel are well aligned to be \emph{on-the-grid} with respect to the same fine grid $\Gamma$.
			Therefore, in contrast to the OTFS, whose DD channel input-output relation (\ref{otfsiorelation}) is an approximation\cite{8424569} due to the OOBE and the off-the-grid TF domain ISI and ICI,  we can achieve an elegant and exact DD channel input-output relation for ODDM, evidenced by $\mathbf H$ in (\ref{H}). In other words, with the PR condition in (\ref{orth}),
			the underlying signal structure of the ODDM modulation perfectly couples with the DD channel by performing orthogonal MC modulation on the DD plane.}
\end{remark}

\begin{figure}
	\centering
	\includegraphics[width=9cm]{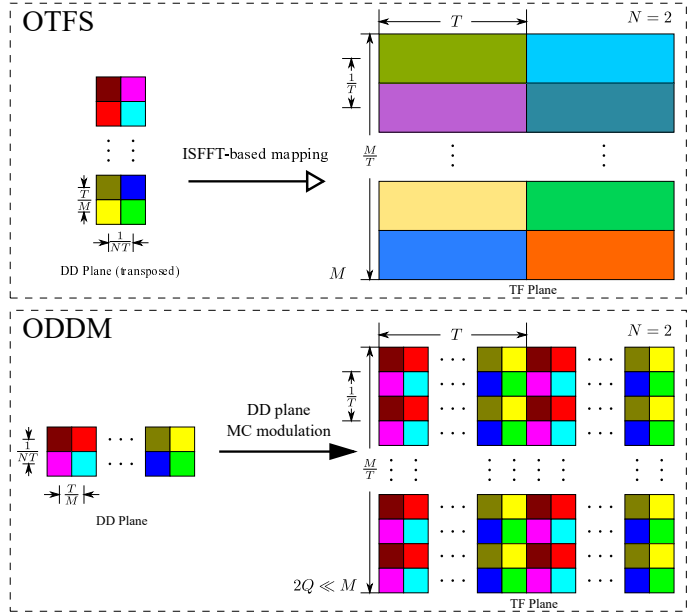}
	\caption{ODDM versus OTFS, $N=2$, $2Q \ll M$.}
	\label{oddmvsotfs}
\end{figure}

The difference between ODDM and OTFS can be clearly observed via a comparison of TF plane signal localization in Fig. \ref{oddmvsotfs}. For OTFS, the DD domain signal is mapped to the TF domain using the ISFFT, therefore its signal localization on the TF plane is just like that of the conventional OFDM. As a result, without an ideal pulse, the OTFS suffers from the blurred TF domain ISI and ICI induced by the doubly-selective channel. On the other hand, by employing $u(t)$, the ODDM staggers $M$ symbols with an interval of $\frac{T}{M}$, inside which the $N$ information-bearing subcarriers with a subcarrier spacing of $\frac{1}{NT}$ are spread $M$ times. Therefore, the ODDM is a hybrid of time division multiplexing and frequency division multiplexing. For a large enough $M\gg 2Q$, any two adjacent pulses (i.e. $a(t)$ and $a(t-T)$) in $u(t)$ become far away and then separated. We can observe that as opposed to the ISFFT-based mapping in OTFS, the ODDM basically increases the TF plane's time and frequency resolutions by $M$ and $N$ times, respectively, to remove the resolution mismatch between the TF and DD planes. Then, the DD domain signals are modulated {on} the densified TF plane in a 2D uniformly distributed fashion, achieving orthogonality with respect to the DD plane's fine resolutions. In other words, \emph{the TF plane is oversampled to a DD plane to perform the ODDM modulation}. As a result, the ODDM enjoys the prefect coupling between the modulated signal and the DD channel, and it only experiences the well-controlled on-the-grid ISI and ICI. Meanwhile, we can see that each pulse $u(t)$ occupies an area not less than $1$ on the TF plane to guarantee its realizability without violating the Heisenberg's uncertainty principle.

\begin{remark}
	Although the proposed ODDM symbol was introduced in (\ref{xm}) based on the sample-wise pulse shaping using $a(t)$,
	{it is the PS-OFDM symbol in (\ref{dotxm}), which is generated using the DD plane orthogonal pulse $u(t)$ and has a DD plane MC modulation form.}
	In other words, an ODDM symbol is essentially a PS-OFDM symbol obtained via the pulse shaping in Fig. \ref{ps_ofdm} with $i(t)=\sinc\left(\frac{t}{T}\right)$ and $g_{tx}(t)=u(t)$.
	When $2Q \ll M$, as proved in Appendix A, the ODDM can be implemented approximately by the pulse shaping using $a(t)$ in (\ref{xm}), which is equivalent to the $a(t)$-based wideband filtering explained in Remark 2.
	Consequently, considering that the ODDM and the OTFS have the same time domain digital sequence, when $2Q \ll M$, a digital OTFS signal filtered by $a(t)$ approximates an ODDM waveform.
\end{remark}

\begin{table}[t]
	\centering
	\caption{Simulation Parameters}
	\begin{tabular}{|c|c|}
		\hline
		Parameter           & Value       \\
		\hline
		Carrier frequency   & 5 GHz       \\
		\hline
		Subcarrier spacing $\frac{1}{T}$  & 15 kHz      \\
		\hline
		$M$                 & 512         \\
		\hline
		$N$                 & 32, 64      \\
		\hline
		CP length           & 3.125$\mu$s \\
		\hline
		Modulation alphabet & 4-QAM       \\
		\hline
		UE speed (km/h)     & 120, 500    \\
		\hline
	\end{tabular}
\end{table}

\subsection{Signal detection}
From (\ref{ioddcompact}) and (\ref{H}), it is clear that ODDM symbols experience interference and an effective data detector is required to unlock the full time and frequency diversity potentials offered by ODDM in order to obtain reliable error performance. The maximum likelihood and maximum a posteriori detector are impractical as they have prohibitive complexity, exponential with the block length $MN$. In the literature, various low complexity detectors have been proposed for the OTFS modulation, including two-stage detector\cite{ref7}, an iterative receiver with minimum mean squared error equalization and parallel interference cancellation with a soft-output sphere decoder\cite{ref8}, message passing techniques\cite{8424569,ref12}, a variational Bayes based detector\cite{ref13}, an iterative rake decision feedback detector\cite{ref10}, the approximate message passing  algorithm\cite{ampotfs}, hybrid or cross-domain iterative detectors\cite{hybrid}\cite{crossdomain}, etc. Considering that the main focus of the paper is a novel DD plane MC scheme, we evaluate the proposed ODDM's performance with a commonly deployed DD domain messaging passing detector, like that for the OTFS modulation. By exploiting the sparsity of the channel matrix in the DD domain, the message passing detector exchanges message/information between observation nodes and variable nodes of the factor graph for the ODDM system. In the message passing detection, we assume that the total interference and noise at each observation node follows a Gaussian distribution.
Its computational complexity is in the order of $MNS$, where $S=P$ is the number of nonzero entries in each row of the DD domain channel matrix.

\section{Simulation Results}
In this section, simulations are conducted to verify the performance of the proposed ODDM modulation. Especially, the ODDM modulation is compared to the OTFS modulation. The simulation parameters are shown in Table I.

For the doubly-selective channel, similar to \cite{8424569}, we adopt the Extended Vehicular A (EVA) model \cite{eva_channel_model}, where each path has a Clarke
Doppler power spectral density (PSD) according to the velocity of the user equipment (UE).
It is noteworthy that the EVA channel has not only off-the-grid channel taps on the delay axis, but also possible off-the-grid Dopplers. Therefore, the channel taps spread over the grid of the DD plane. In other words, these spread on-the-grid channel taps represent the equivalent DD channel of the EVA channel. Meanwhile, for the ODDM modulation, a square-root raised cosine pulse with a roll-off factor of $0.1$ is employed as $a(t)$, where $Q=20$. Also, the interpolation filter $\sinc\left(\frac{t}{T}\right)$ is truncated at its $50$th zero-crossing point. The bandwidth of the ODDM is therefore about $8.45$ MHz.

\begin{figure}
	\centering
	\includegraphics[width=8.5cm]{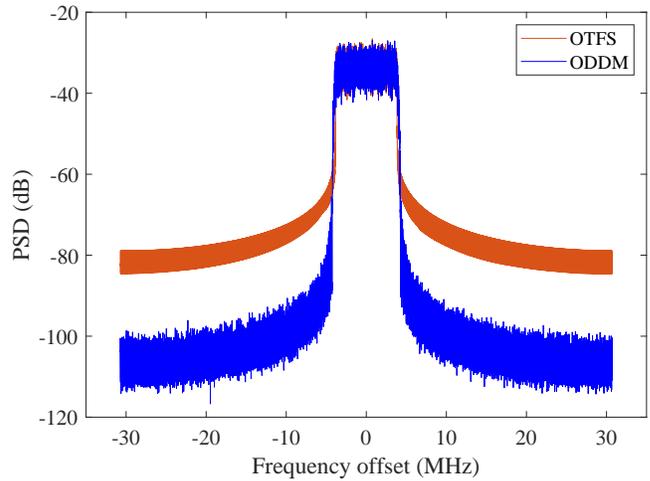}
	\caption{PSD comparison, $M=512$, $N=64$, $4$-QAM.}
	\label{powspec_comp}
\end{figure}

The PSD comparison of the modulated signals is shown in Fig. \ref{powspec_comp}. There is no surprise to find that because of the square-root Nyquist pulses based shaping, the ODDM modulation has much lower OOBE than the OTFS modulation, and up to $20$ dB improvement can be observed at the expense of excess bandwidth. {Note that due to the severe OOBE, the bandwidth of a fully-loaded OFDM system is not well-defined. Therefore, in order to sharpen the spectrum, a practical OFDM system does not fully utilize all subchannels}, leading to a reduced spectral efficiency considering some null-subcarriers placed at the band edge.
This fact also applies to the OTFS based on the TF plane OFDM interpretation, which however may become an issue. Because the information-bearing symbols are modulated in the DD domain, it is still unclear that how to arrange them in the DD domain to achieve unloaded edge subcarriers in the TF domain after the ISFFT, see the comparisons of TF plane signal distribution in Fig. \ref{oddmvsotfs}. A possible solution is to reduce $M$, which however is the number of subcarriers that is usually a power of $2$ in practice to exploit low-complexity fast Fourier transform (FFT) and therefore cannot be changed freely. On the other hand, the proposed novel ODDM scheme can fully utilize the number of subchannels. By tuning the roll-off factor, a trade-off between the excess bandwidth and OOBE can be struck to achieve the desirable spectral efficiency. Furthermore, since $M$ now is the number of ODDM symbols rather than the number of subcarriers, {it is not a necessity to be a power of $2$} and therefore can be chosen flexibly to adjust the bandwidth together with the roll-off factor.

\begin{figure}
	\centering
	\includegraphics[width=8.5cm]{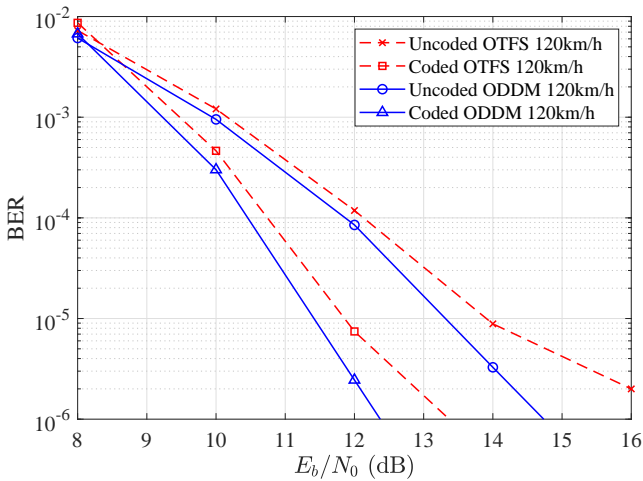}
	\caption{BER comparison, $M=512$, $N=64$, $4$-QAM.}
	\label{ber_comp_512_64}
\end{figure}

\begin{figure}
	\centering
	\includegraphics[width=8.5cm]{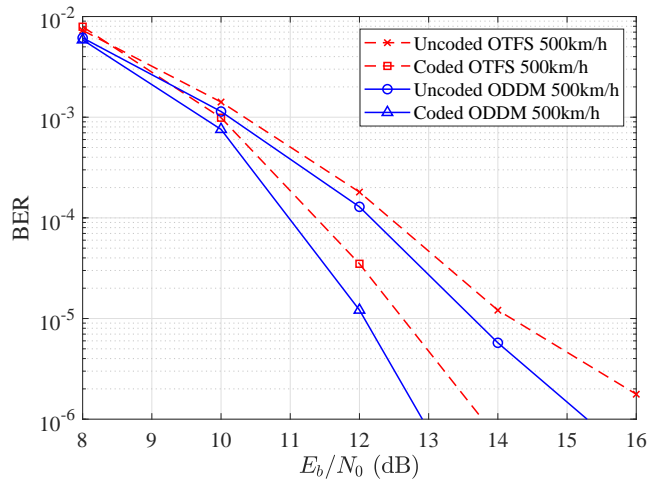}
	\caption{BER comparison, $M=512$, $N=32$, $4$-QAM.}
	\label{ber_comp_512_32}
\end{figure}

We now evaluate the BER performance of the uncoded and coded ODDM and OTFS modulations.
For the ODDM modulation, the signal detection is based on the message passing algorithm and the DD domain channel matrix $\mathbf H$ in (\ref{H}).
Figure \ref{ber_comp_512_64} shows the BER of the proposed ODDM system with $4$-QAM signals, $M=512$, $N=64$, and UE speed of $120$ km/h. For comparison, we also show the BER performance of the OTFS system with message passing detection. For the coded systems, we employ a rate $2/3$ convolutional code with constraint length $5$ and generator polynomials of $[23, 35, 0; 0, 5, 13]$ in octal and the Viterbi decoding algorithm is used at the receiver.
It can be seen clearly from Fig. \ref{ber_comp_512_64} that the uncoded ODDM outperforms the uncoded OTFS system by about $1.7$ dB at the BER of $ 2\times 10^{-6}$. We also observe that about $0.8$ dB  performance gain is achieved for the coded ODDM system over the coded OTFS system at the BER of $ 2\times 10^{-6}$.
The error performance of the proposed ODDM system with $4$-QAM signals, $M=512$, $N=32$, and UE speed of $500$ km/h is plotted in Fig. \ref{ber_comp_512_32}. In this case, we can observe from Fig. \ref{ber_comp_512_32} that the uncoded and coded ODDM outperform the OTFS counterpart by about $1.1$ dB and $0.8$ dB, respectively, at the BER of $2\times 10^{-6}$.
	{Figures \ref{ber_comp_512_64} and \ref{ber_comp_512_32} demonstrate that ODDM can achieve robust performance gain over OTFS for high mobility channels with different Doppler resolutions. Note that different parameter settings, for example, $M$, $N$, channel profile, detection algorithm, etc, lead to different performance gains of ODDM over OTFS.}

We would like to point out that due to the orthogonality of the transmit pulse $u(t)$ that fulfills the PR condition with respect to the DD resolutions, the ODDM not only uses matched filters as receive filters to maximize the signal-to-noise ratio, but also has an exact DD domain channel input-output relation, which can be exploited by the receiver to accurately detect the transmitted signals. On the other hand, the OTFS's DD domain channel input-output relation is just an approximation due to the complicated TF domain ISI and ICI. Therefore, both the matched filtering and exact input-output relation of the ODDM contribute to its better performance than the OTFS.

	{In this section, we evaluated uncoded and coded ODDM performance. In practical systems,  other issues such as spectral mask and synchronization errors, should also be considered for performance evaluation. In the future, we will investigate these issues and compare the ODDM with other OTFS or precoded-OFDM systems, such as those in \cite{zemenpimrc2018}.}

\section{Conclusion}
We studied the problem of MC modulation on DD plane, and revealed the link between the DD plane MC modulation and the conventional SMT modulation. It was pointed out that staggering MC symbols is the key to find a realizable orthogonal pulse for DD plane MC modulations. We then proposed the ODDM modulation, whose staggered upsampled-OFDM representation in digital domain was presented. Based on the spectrum analysis, a sample-wise square-root Nyquist pulse shaping was employed to approximately generate the analog ODDM waveform. The transmit pulse of the ODDM was identified, and its orthogonality with respect to the DD  plane's resolutions was proved. By virtue of this favorable orthogonality, we derived an exact DD domain channel input-output relation of the proposed ODDM, where the DD domain channel matrix has an elegant block-circular-like structure. Finally, thanks to the perfect coupling between the modulated signal and the DD channel, the ODDM presents the superior performance over the OTFS in terms of OOBE and BER, which was demonstrated by simulations.  The exact DD domain channel input-output relation lays a foundation to design efficient detection algorithms for the ODDM in the future.


%
\section{Acknowledgement}
We thanks the editor and the reviewers for their constructive comments.

\appendices
\section{Proof of $\check g_{tx}(t)\approx u(t)$ for ODDM generated using $a(t)$}
Since $2Q \ll M$, most portions of $u(t)$ is zero. The difference between $x_m(t)$ in (\ref{xm}) and $\tilde x_m(t)$ in (\ref{dotxm}) only exists in the portions of $\left(\dot nT-\frac{QT}{M},\dot nT+\frac{QT}{M}\right)$ for $\dot n=0,\ldots, N-1$. Comparing (\ref{xm}) to (\ref{dotxm}),  it can be found that $x_m(t)=\tilde x_m(t)$ when $t=\dot nT$, for $\dot n=0,\ldots, N-1$.
Let $\dot \tau =t-\dot n T$, the difference between $x_m(t)$ and $\tilde x_m(t)$ within the range $\left(\dot nT-\frac{QT}{M},\dot nT+\frac{QT}{M}\right)$ is
\begin{equation} \nonumber
	e(\dot \tau)= \left(\sum_{n=0}^{N-1} X(m,n) e^{\frac{j2\pi n\dot n }{N}} - \sum_{n=0}^{N-1} X(m,n) e^{\frac{j2\pi n \dot n }{N}}e^{\frac{j2\pi n \dot \tau}{NT}}\right) \times a(\dot \tau),
\end{equation}
for $\dot \tau \in \left(-\frac{QT}{M},\frac{QT}{M}\right)$.
	{Since
		$\frac{2\pi n \dot \tau}{NT} \in \left( \frac{-2\pi n Q}{MN},\frac{2\pi n Q}{MN} \right)$ and
		$2Q \ll M$, we have $e^{j\frac{2\pi n \dot \tau}{NT}} \approx 1$ then $e(\dot \tau)\approx 0$. The approximation error of $e^{j\frac{2\pi n \dot \tau}{NT}} \approx 1$,} although very small, increases as $|\dot \tau|$ increases. Meanwhile, since $|a(\dot \tau)|$ decrease as $|\dot \tau|$ increases, the overall approximation error of $e(\dot \tau)\approx 0$ is greatly reduced. Consequently, $e(\dot \tau)$ is negligibly small, and we have $x_m(t)\approx \tilde x_m(t)$, which completes the proof.

\section{Proof of the orthogonality of $u(t)$}
Since the real-valued filter $a(t)$ only has support on $\left\{-\frac{QT}{M}, \frac{QT}{M}\right\}$ and $2Q \ll M$, for $|m| \le M-2Q$,
the ambiguity function of $u(t)$ is given by
\begin{align}
	 & A_{u,u}\left(m\frac{T}{M},n\frac{1}{NT}\right)
	=  \int u(t)u\left(t-m\frac{T}{M}\right)e^{-j\frac{2\pi n}{NT}(t-m\frac{T}{M})} dt, \nonumber                                                                                                                                              \\
	 & =  \sum_{\dot n=0}^{N-1} \int_{\dot nT-\frac{QT}{M}}^{\dot nT+\frac{QT}{M}} a(t-\dot nT)a\left(t-\dot nT-m\frac{T}{M}\right) e^{-j\frac{2\pi n}{NT}(t-m\frac{T}{M})} dt, \label{au1}                                                    \\ %
	 & =  \sum_{\dot n=0}^{N-1} \int_{-\frac{QT}{M}}^{\frac{QT}{M}} a(\dot \tau)a\left(\dot \tau-m\frac{T}{M}\right)e^{-j\frac{2\pi n }{NT}(\dot \tau-m\frac{T}{M})}  e^{-j\frac{2\pi n \dot n T}{NT}} d\dot \tau, \nonumber                   \\
	 & =  \sum_{\dot n=0}^{N-1} e^{-j\frac{2\pi n \dot n }{N}}\times \int_{-\frac{QT}{M}}^{\frac{QT}{M}} a(\dot \tau)a\left(\dot \tau-m\frac{T}{M}\right)  e^{-j\frac{2\pi n }{NT}(\dot \tau-m\frac{T}{M})} d\dot \tau, \nonumber              \\
	 & =  e^{j\frac{2\pi m n }{MN}}\sum_{\dot n=0}^{N-1} e^{-j\frac{2\pi n \dot n }{N}}\times \int_{-\frac{QT}{M}}^{\frac{QT}{M}} a(\dot \tau)a\left(\dot \tau-m\frac{T}{M}\right)  e^{-j\frac{2\pi n \dot \tau}{NT}} d\dot \tau . \label{au2}
\end{align}
Bearing in mind that $\displaystyle{\sum_{\dot n=0}^{N-1} e^{-j\frac{2\pi n \dot n }{N}} =0} $ for $n\ne 0$ and $a(t)$ is a square-root Nyquist pulse for the symbol period of $\frac{T}{M}$, (\ref{au2}) becomes
\begin{eqnarray}\label{au3}
	A_{u,u}\left(m\frac{T}{M},n\frac{1}{NT}\right) & = & N \delta(n) \int_{-\frac{QT}{M}}^{\frac{QT}{M}} a(\dot \tau)a\left(\dot \tau-m\frac{T}{M}\right) d\dot \tau \nonumber, \\
	& = & \delta(m) \delta(n).
\end{eqnarray}
When $M-2Q<m\le M-1$, we can let $\dot m= m-M$. Then, notice that $0< |\dot m|<2Q\le M-2Q$, like (\ref{au1}), we have
\begin{align}
	 & A_{u,u}\left(m\frac{T}{M},n\frac{1}{NT}\right) = \int u(t)u\left(t-m\frac{T}{M}\right)e^{-j\frac{2\pi n}{NT}(t-m\frac{T}{M})} dt, \nonumber                                                                                                     \\
	 & \stackrel{(a)}{=} \sum_{\dot n=1}^{N-1} \int_{\dot nT-\frac{QT}{M}}^{\dot nT+\frac{QT}{M}} a(t-\dot nT)a\left(t+T-\dot nT-m\frac{T}{M}\right) e^{-j\frac{2\pi n}{NT}(t+T-m\frac{T}{M})} dt, \nonumber                                           \\
	 & = \sum_{\dot n=1}^{N-1} \int_{\dot nT-\frac{QT}{M}}^{\dot nT+\frac{QT}{M}} a(t-\dot nT)a\left(t-\dot nT-\dot m\frac{T}{M}\right) e^{-j\frac{2\pi n}{NT}(t-\dot m\frac{T}{M})} dt, \nonumber                                                     \\
	 & = e^{j\frac{2\pi \dot m n }{MN}}\sum_{\dot n=1}^{N-1} e^{-j\frac{2\pi n \dot n }{N}}\times \int_{-\frac{QT}{M}}^{\frac{QT}{M}} a(\dot \tau)a\left(\dot \tau-\dot m\frac{T}{M}\right)  e^{-j\frac{2\pi n \dot \tau}{NT}} d\dot \tau, \label{au4}
\end{align}
where $\stackrel{(a)}{=}$ is due to the fact that the pulses $a(t-\dot nT)$ and $a(t-\dot nT-m \frac{T}{M})$ do not overlap but $a\left (t-\dot nT\right)$ and $a\left(t+T-\dot nT-m \frac{T}{M}\right) $ overlap for these $m$ values. When $n=0$, (\ref{au4}) becomes
\begin{eqnarray*}
	A_{u,u}\left(m\frac{T}{M},n\frac{1}{NT}\right)  =  (N-1) \times \int_{-\frac{QT}{M}}^{\frac{QT}{M}} a(\dot \tau)a\left(\dot \tau-\dot m\frac{T}{M}\right)  d\dot \tau = 0.
\end{eqnarray*}
When $n\ne 0$, from Appendix A, we know that $e^{\frac{j2\pi n \dot \tau}{NT}} \approx 1$ for $\dot \tau \in \left(-\frac{QT}{M},\frac{QT}{M}\right)$. Bearing in mind that $\sum_{\dot n=1}^{N-1} e^{-j\frac{2\pi n \dot n }{N}}=-1$ for $n\ne 0$, (\ref{au4}) then becomes
\begin{eqnarray*}
	A_{u,u}\left(m\frac{T}{M},n\frac{1}{NT}\right) \approx  -e^{j\frac{2\pi \dot m n }{MN}}\times \int_{-\frac{QT}{M}}^{\frac{QT}{M}} a(\dot \tau)a\left(\dot \tau-\dot m\frac{T}{M}\right)  d\dot \tau =  0,
\end{eqnarray*}
where the approximation error is negligibly small, similar to that in Appendix A. Meanwhile, for $-M+1 \le m <-M+2Q$, we can have results similar to (\ref{au4}), with different $\dot m$ and the corresponding different range for the summation indexed by $\dot n$. The combination of (\ref{au3}) and (\ref{au4}) completes the proof.

\ifCLASSOPTIONcaptionsoff
	\newpage
\fi




\bibliographystyle{IEEEtran}
\bibliography{oddm}
\end{document}